\documentclass{iau} 
\usepackage{graphicx}
\usepackage{xcolor}
\usepackage{xparse}
\usepackage{amsmath}
\usepackage{bm,upgreek}
\usepackage[authoryear]{natbib}
\usepackage{hyperref}

\newcommand{\HI}{H\textsc{i}}
\newcommand{\kpar}{\ensuremath{k_{||}}}

\ExplSyntaxOn
\NewDocumentCommand\vect{m}
{
	\commexo_vector:n { #1 }
}

\cs_new_protected:Npn \commexo_vector:n #1
{
	\tl_map_inline:nn { #1 }
	{
		\commexo_vector_inner:n { ##1 }
	}
}

\cs_new_protected:Npn \commexo_vector_inner:n #1
{
	\tl_if_in:VnTF \g_commexo_latin_tl { #1 }
	{
		\mathbf { #1 } 
	}
	{
		\tl_if_in:VnTF \g_commexo_ucgreek_tl { #1 }
		{
			\bm { #1 } 
		}
		{
			\tl_if_in:VnTF \g_commexo_lcgreek_tl { #1 }
			{
				\commexo_makeboldupright:n { #1 }
			}
			{
				#1 
			}
		}
	}
}

\cs_new_protected:Npn \commexo_makeboldupright:n #1
{
	\bm { \use:c { up \cs_to_str:N #1 } }
}

\tl_new:N \g_commexo_latin_tl
\tl_new:N \g_commexo_ucgreek_tl
\tl_new:N \g_commexo_lcgreek_tl
\tl_gset:Nn \g_commexo_latin_tl
{
	ABCDEFGHIJKLMNOPQRSTUVWXYZ
	abcdefghijklmnopqrstuvwxyz
}
\tl_gset:Nn \g_commexo_ucgreek_tl
{
	\Gamma\Delta\Theta\Lambda\Pi\Sigma\Upsilon\Phi\Chi\Psi\Omega
}
\tl_gset:Nn \g_commexo_lcgreek_tl
{
	\alpha\beta\gamma\delta\epsilon\zeta\eta\theta\iota\kappa
	\lambda\mu\nu\xi\pi\rho\sigma\tau\upsilon\phi\chi\psi\omega
	\varepsilon\vartheta\varpi\varphi\varsigma\varrho
}

\ExplSyntaxOff

\title[Clustered Point-Source Foregrounds in EoR] 
{A Clustered Extragalactic Foreground Model for the EoR}

\author[Steven G. Murray]   
{S.~G.~Murray$^{1,2}$,
 C.~M.~Trott$^{1,2}$
 \and C.~H.~Jordan$^{1,2}$
 }

\affiliation{$^1$ARC Centre of Excellence for All-sky Astrophysics (CAASTRO) \\[\affilskip]
$^2$International Centre for Radio Astronomy Research (ICRAR), Curtin University,  Bentley, WA 6102, Australia}

\pubyear{2018}
\volume{333}  
\jname{Peering towards Cosmic Dawn}
\editors{Vibor Jeli\'c \& Thijs van der Hulst, eds.}
\begin{document}

\maketitle

\begin{abstract}
We review an improved statistical model of extra-galactic point-source foregrounds first introduced in \citet{Murray2017}, in the context of the Epoch of Reionization.
This model extends the instrumentally-convolved foreground covariance used in inverse-covariance foreground mitigation schemes, by considering the cosmological clustering of the sources. 
In this short work, we show that over scales of $k \sim (0.6, 40.) h {\rm Mpc}^{-1}$, ignoring source clustering is a valid approximation.
This is in contrast to \citet{Murray2017}, who found a possibility of false detection if the clustering was ignored. 
The dominant cause for this change is the introduction of a Galactic synchrotron component which shadows the clustering of sources.
\end{abstract}

\firstsection 
\section{Introduction}
The power spectrum (PS) of the brightness temperature fluctuations in the redshifted 21~cm emission of neutral hydrogen (\HI) remains the most promising probe of the periods of first light -- the Cosmic Dawn, and Epoch of Reionization (EoR). 
Indeed, several low-frequency interferometers are yielding ever-tightening upper-limits on the amplitude of the PS at a range of redshifts and physical scales (eg. \citet{Ali2015}, \citet{Beardsley2016}, \citet{Patil2017}).

The outstanding obstruction in these efforts is the presence of overwhelming systematics, arising from both the instrument and foreground emission, as well as the complex interaction of these factors.
Many studies have investigated means of mitigating foregrounds.
Each method has a number of advantages and drawbacks, but in this work we focus on foreground \textit{suppression}, as espoused by the \textsc{chips}\footnote{Cosmological \HI\ Power-Spectrum Estimator} \citep[hereafter T16]{Trott2016} PS code -- one of two primary estimators used within the MWA project. 

The essence of foreground suppression lies in noting that while we typically consider the EoR signal to be three-dimensionally isotropic, and therefore able to be averaged in spherical shells, the foregrounds are decidedly not. 
In particular, foregrounds have drastically different structure along the line of sight -- scales probed by frequency -- than they do on the sky. 
Indeed, line-of-sight foreground structures are expectedly smooth, and consequently their power appears almost exclusively at low Fourier modes, $\kpar$ (with the exception that instrumental effects cause mode-mixing, resulting in the famous ``wedge").
Foreground suppression merely assumes that we may model the expected contribution to, and uncertainty of, the PS at any given 2D mode, and consistently weight each mode when averaging to the final 1D PS.


The determining factor of the foreground suppression scheme is the accuracy of its covariance. 
The \textsc{chips} framework favours a parametric approach, in which each foreground source -- compact extra-galactic sources, Galactic synchrotron or free-free -- is statistically defined to produce a \textit{model} covariance.
Thus a great deal hinges on the complexity of the constituent statistical models.
In this work, we revisit one of the simplifying assumptions made in the original \textsc{chips} framework, namely that of uniform angular distribution of extra-galactic point-sources, and derive an extension to isotropically correlated structure. 

This brief overview will first present the  model in \S\ref{sec:derivation} before illustrating its implications in \S\ref{sec:results}.
We refer the interested reader to \citet[hereafter M17]{Murray2017} for a fuller treatment of the work presented here.

\section{Model Covariance}
\label{sec:derivation}
In this work, we exclusively deal with the covariance arising from extra-galactic point-sources which are not peeled during calibration (typically because they are too faint to individually model). 
We will first present the model (along with its assumptions) introduced by the \textsc{chips} framework, which will then be conceptually simple to extend. 

\subsection{\textsc{chips} Covariance Model}
The \textsc{chips} extra-galactic point-source model covariance, denoted here as $\vect{C}_{\rm unif}$, requires three basic statistical properties of the source population: (i) the source counts as a function of their flux density, (ii) their average spectral energy distribution (SED) and (iii) their distribution on the sky. The source counts are modelled by a single power-law, which is truncated at low ($S_{\rm min}$) and high ($S_{\rm max}$) flux density to simulate the effects of a luminosity function turnover and peeling limit respectively. The SED is taken to be a power law, with a universal slope of $\gamma = 0.8$. Finally, the angular distribution is taken to be uniform, with counts in infinitesimal bins drawn from a Poisson distribution. This last model is the assumption we most heavily revise in this work. 

Along with these models, the properties of the hypothetical interferometer must be specified. 
In this work we make the simplifying assumption of an ideal instrument which samples all spatial scales equally and suffers no channelisation issues. 
Furthermore, for all examples we will use a simple frequency-dependent Gaussian beam:
\begin{equation}
B(l) = e^{-l^2/2\sigma_\nu^2}, \ \ \ \sigma_\nu \propto \frac{1}{\nu D},
\end{equation}
with $l$ the sine of the angle from zenith, and $D$ the diameter of the instrument's tiles.

With these simplistic models, we are able to determine the covariance between visibilities on the \textit{same baseline} at \textit{different frequencies}:
\begin{align}
\vect{C}_{\rm unif}(f', f'', \vect{u}) = 2\pi (f' f'')^{-\gamma}  \mu_2 \int_0^\infty  l B'B'' J_0(2\pi u l)  \ dl.
\end{align}
Here $f'$ and $f''$ are the frequencies of the covarying visibilities, normalised by some reference frequency, $\nu_0$, $\vect{u}$ is the Fourier-dual of $\vect{l}$ and $J_0$ is the zeroth-order Bessel function of the first kind. 
We have left the integral over the beam unevaluated for generality, but this may be determined analytically for the simple frequency-dependent Gaussian beam defined above.
Finally, $\mu_2$ is the second moment of the source-count distribution:
\begin{equation}
	\label{eq:mu}
	\mu_n = \int S^n \frac{dN}{dS}dS \ \ \ [{\rm Jy}^n {\rm sr}^{-1}].
\end{equation}


\subsection{Extended Covariance Model}

Our extension is to the spatial distribution, which we approximate with isotropically correlated structure, defined by a source power spectrum, $P_{\rm src}(u)$, which we here assume to be a power-law: $P_{\rm src} = (u/u_0)^{-\kappa}$.
Under the assumption of Gaussian density fluctuations (which is valid for the integrated source counts), we find that, over all scales of interest, the new covariance is 
\begin{equation}
\label{eq:solution}
\vect{C}(f', f'', u) =  \vect{C}_{\rm unif} + y\mu_1^2 Q_\nu P_{\rm src}(u) e^{\pi y u^2 (q-2)},
\end{equation}
where $q = (1+p)^2/(1+p^2)$ (with $p=f'/f''$) has a maximum of 2 when $f' \equiv f''$, and
\begin{equation}
	Q_\nu = \frac{q^{-\kappa/2}(1+p^2)^{\kappa/2-1}}{{f'}^{\kappa+\gamma}{f''}^{2+\gamma}}.
\end{equation}

\section{Results}
\label{sec:results}

\subsection{Regime of Clustering Dominance}
While the full covariance exhibits a rather complex dependence on the coupled frequencies, the most important aspect is the variance, where $f' = f''$. 
Within this subset, we may directly calculate at which scales $u$ the clustering term of the covariance dominates.
For the natural choice of a negatively-sloped $P_{\rm src}$, this will be at the largest scales, and we can define a crossover scale $u_\star$ which is given by
\begin{equation}
\label{eq:ustar}
u_\star = \left(\frac{\mu_1^2}{\mu_2}\right)^{1/\kappa}  u_0 .
\end{equation}
This has a dependence on the source count distribution through $\mu_1, \mu_2$, and also on $P_{\rm src}$ through $\kappa$ and $u_0$.
Interestingly, the source-count dependence indicates that $u_\star$ grows as the relative fraction of faint sources in the population increases. 
Thus, while the overall amplitude of foreground noise will decrease for future SKA observations (through more aggressive peeling), the relative importance of taking into account the clustering term as presented here will increase.

\subsection{PS Estimation with Improved Model}
To determine the effect of ignoring the clustering of foregrounds on estimation of the PS, we use data from the \textit{Evolution of Structure} (EoS) project (specifically the \textsc{faint galaxies} simulation) \citep{Mesinger2010}.
We add foreground point sources according to the model presented here (for fiducial parameters, refer to Table 1 of M17), and neglect thermal noise.

In addition to the pure point-source model of M17, in this work we add a simple model of Galactic synchrotron, which is modelled by a spatial power-law power spectrum with mean brightness and fluctuation level given by a quiet patch of the Global Sky Model\footnote{As implemented by \textsc{PyGSM}, \url{https://github.com/telegraphic/PyGSM}} \citep{Zheng2017}. Explicitly, the covariance for this model is given by the second term of Eq. \ref{eq:solution}, with $\mu_1$ replaced by $\bar{S} = 0.13$ Jy and $P_{\rm src}$ replaced by $P_{\rm gal} = (u/0.103)^{-2.7}$, along with a spectral index of 0.55. 



\begin{figure}
\centering
\includegraphics[width=\linewidth,trim=0cm 0cm 1cm 0cm]{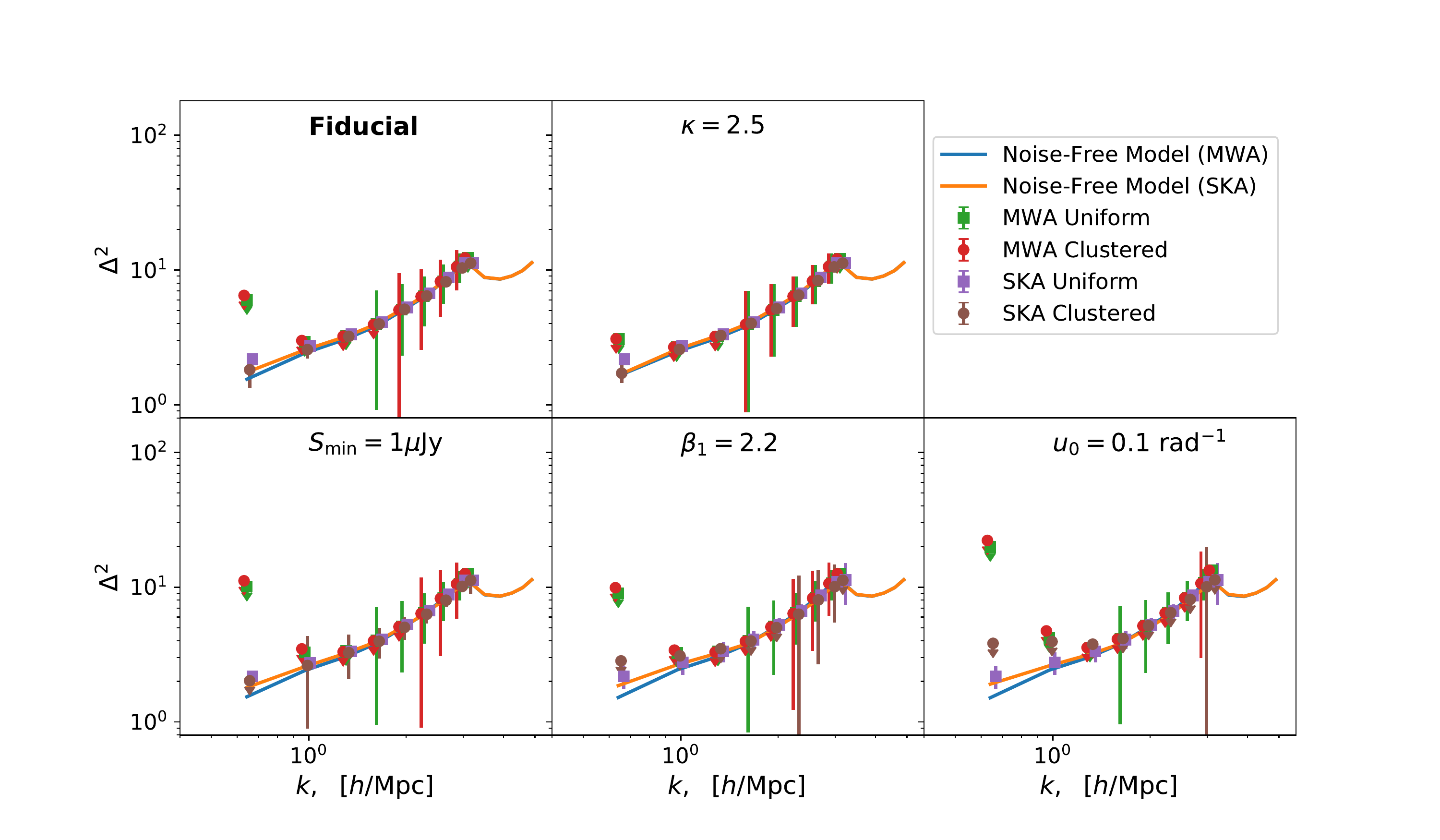}
\caption{1D PS estimates for various foreground model instances (different panels). In each panel, the true power is shown as solid lines (where each colour represents different intra-bin weighting). Circles show estimates ignoring source clustering, while squares incorporate the clustering model (which is present in the foregrounds), and both are shown for an MWA-like and SKA-like observation. Upper limits are indicated with short arrows.}
\label{fig:final_1d}
\end{figure}

Figure \ref{fig:final_1d} shows the resulting spherically averaged 1D PS for a range of statistical foreground model instances.
In addition to the inclusion of a Galactic component, two major changes are present in Figure \ref{fig:final_1d} as compared to Figure 10 of M17: (i) The ``noise-free'' model is here shown as a weighted average, using the statistical covariance model weights, and (ii) a more accurate estimate of the foreground-contributed power is made. The first change serves to alleviate the visible bias that arises from the evolution of the signal within a $k$-bin. The second change uses the variance of the visibility as the estimate rather than the square root of the covariance of the power. The effect is to decrease the total power, and therefore the bias, of all estimates.

Due to these modifications, especially the incorporation of a (perfectly known) Galactic component, the estimates shown here do not suffer from the possibility of ``false detection'', in the sense that ignoring clustering does not lead to significantly incorrect estimates, except potentially in the case of the very highest-$k$ mode where $\kappa=2.5$ for the SKA. Despite these results, it will be important to consider the clustering of point-sources in future analyses, as it is the larger scales out of range for this simple simulation that bear the majority of the bias. Furthermore, it is unclear to what extent the layout of baselines will affect these results, and this will be considered rigorously in future work.


\end{document}